# Soft-CP: A Credible and Effective Data Augmentation for Semantic Segmentation of Medical Lesions

Pingping Dai[1], Licong Dong[2], Ruihan Zhang[1], Haiming Zhu[1], Jie Wu[1], Kehong Yuan[1]

[1]Tsinghua University   [2]Peking University Shenzhen Hospital

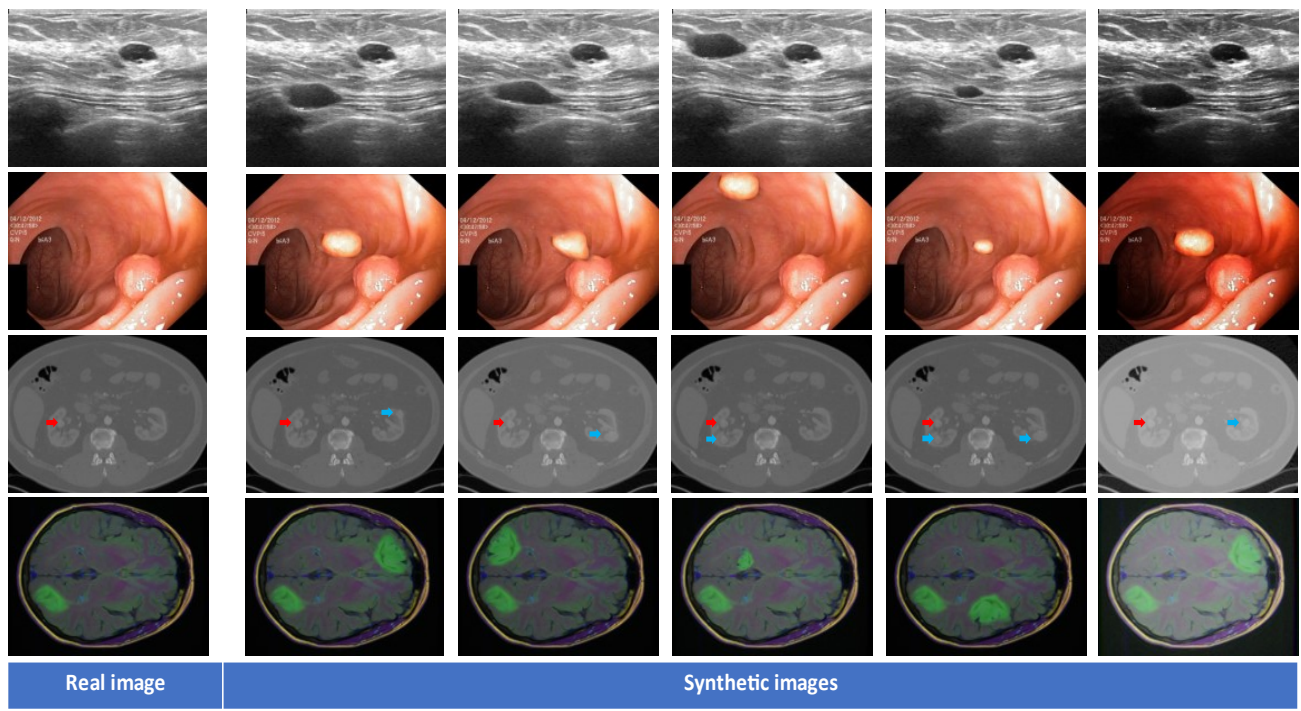

**Figure 1.** Our method, soft copy-paste (short in *Soft-CP*), allows users to augment credible medical lesions for semantic segmentation tasks effectively.

## Abstract

*The medical datasets are usually faced with the problem of scarcity and data imbalance. Moreover, annotating large datasets for semantic segmentation of medical lesions is domain-knowledge and time-consuming.  In this paper, we propose a new object-blend method(short in soft-CP) that combines the Copy-Paste augmentation method for semantic segmentation of medical lesions offline, ensuring the correct edge information around the lession  to solve the issue above-mentioned. We proved the method's validity with several datasets in different imaging modalities. In our experiments on the KiTS19[2] dataset, Soft-CP outperforms existing medical lesions synthesis approaches. The Soft-CP augementation provides gains of +26.5% DSC in the low data regime(10% of data) and +10.2% DSC in the high data regime(all of data), In offline training data, the ratio of  real images to synthetic images is 3:1.*

## 1. Introduction

Semantic segmentation of medical lesions [1,2,3,4,5,6] is an essential and active task in medical imaging diagnosis. In recent years, many effective  medical image segmentation methods based on deep learning have been proposed.e.g, segmentation  frameworks[7,8,9,10,11,12], loss functions[13,14,15,16], training tricks[17,18]. There are also many international competitions1 are held to improve the algorithm for accurate segmentation of various lesions. However, medical images still face data-hungry and data imbalance in deep learning. In contrast, the data augmentation of medical lesions has received little attention[19]. On the one hand, due to the differences between imaging equipment, it is difficult to generalize the medical images collected on one device to other devices. On the other hand, high-quality annotated datasets require intense labor and domain knowledge, consuming enormous time and cost in the medical domain. Furthermore, there are many kinds of human diseases, so that it is difficult for images of rare diseases to form a large public dataset as COCO[20], ADE20K[21], BraTS[1], KiTS[2].

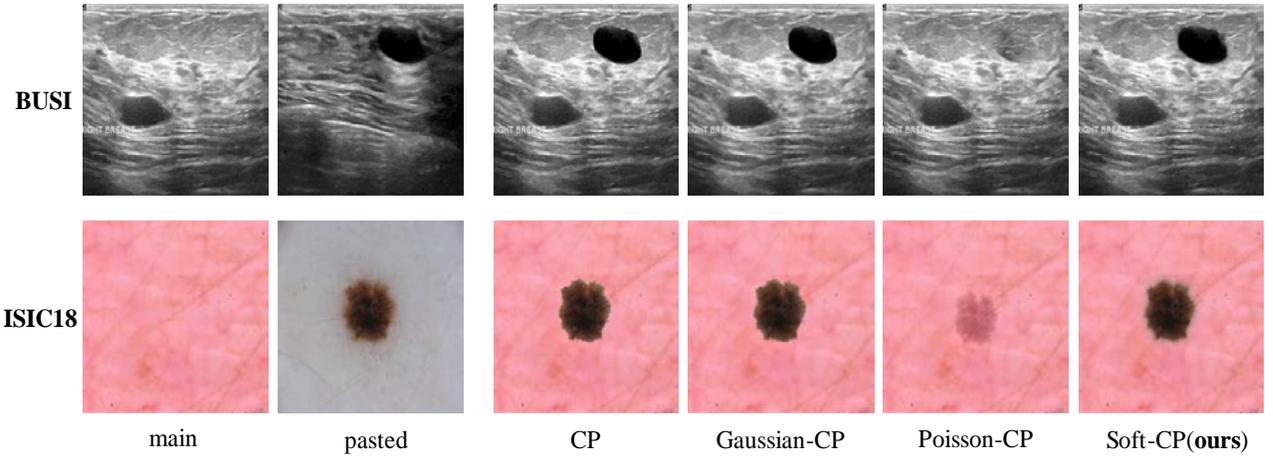

Figure 2. Different blending modes combine Copy-Paste(CP) augmentation method. These modes help lesions paste naturally on main image.

Data augmentation is a direct method to improve data efficiency for semantic segmentation of medical lesions. e.g., rotation, scale jittering, random crop, etc., have been widely used for semantic segmentation. However, they cannot increase the diversity of lesions to solve the problem of data imbalance. Generative Adversarial Networks is another active data augmentation approach[22,23]. However, its good results are based on vast amounts of training data, and the results cannot be controlled well. Mixmatch[29] uses complex operations to decide where to paste crops combined with Mixup augmentation[30]. In contrast, the Copy-Paste augmentation [24,25,26,27,28] is an effectively and simply object-aware method for semantic segmentation by pasting diverse objects to new backgrounds.

However, compared with natural images, medical lesions' surrounding contexts have more professional clinical medical knowledge, so that people remain sceptical about Copy-Paste augmentation's effectiveness for medical segmentation tasks since its surrounding context to be in high requirement. To address this problem, Dwibedi et al.[24] use Poisson blending[64] to smooth the edge of the pasted object. TumorCP[28] uses a Gaussian blur to keep the surrounding context in the medical domain. This method is limited to the imaging modality in which lesions are similar to the background (e.g., Computed Tomography). For imaging modality with large background differences (e.g., Ultrasound, Optical), Gaussian blur is not enough to keep the surrounding context. Intuitively, we show our method(Soft-CP), Poisson blending(Poisson-CP), and Gaussian blur (Gaussian-CP) combined with copy-paste augmentation in Fugure2, Soft-CP blends lesions into the background closer to the real image than Gaussian -CP and Poisson-CP. Moreover, Poisson-CP even changes the lesion itself, which is not what we want. Furthermore, Gaussian -CP augmentation method even has a negative impact on BUSI dataset ( show in table 3).

In this article, we proposed the Soft-CP augmentation method, which has three contributions to the semantic segmentation of medical lesions: i) Guaranteed the correct surrounding context of medical lesions combined with the Copy-Paste augmentation method. ii) Improved visualization and controllability through the offline augmentation method. iii)solved the problem of data imbalance effectively. Furthermore, we conduct experiments on several datasets from different imaging modes, including the BUSI[35], the ISIC18[6], the Kvasir-SEG[5], the KiTS19[2], the LGG[66]. We show that Soft-CP augmentation help baseline(Unet) work better.

## 2. Related Work

### 2.1. Semantic Segmentation of Medical Lesions

Semantic segmentation of medical lesions plays a key role in computer-aided diagnosis in medical imaging due to its accuracy and efficiency. Popular tasks include brain and brain-tumor segmentation[1,31], kidney and kidney-tumor segmentation[2], liver and liver-tumor segmentation [32], polyp segmentation[5,33,34], breast cancer segmentation [35,36], melanoma segmentation[6], lung and pulmonary nodules' segmentation[37], diabetic retinopathy segmenta-tion[38], etc. With the rapid development of deep learning, supervised[7,43,47,48,49], weakly supervised [40,41,44,45], and unsupervised [39,42,46] learning approaches have been widely used for medical image segmentation. So far, supervised learning is still the most popular method since these tasks usually achieve high accuracy, which seriously depends on high-quality labels. Therefore, weakly supervised and unsupervised learning approaches fill the gap by consuming heavily on annotating high-quality labels.

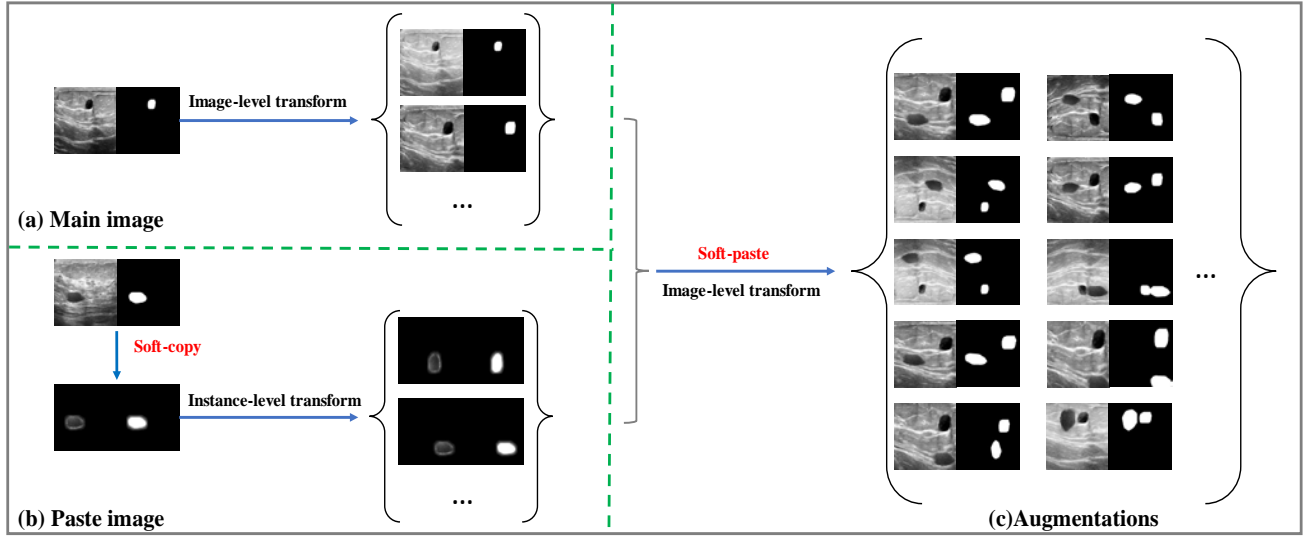

**Figure 3**. Overview of Soft-CP augmentation method proposed in this paper. The main image and paste images are randomly sampled from the dataset. (a) The main image provides several backgrounds after image-level transforming. (b) The lesion provides numerous objects after object-level transformation, copied with "Soft-CP" from the pasted image. (c) The background and object randomly dominate with "Soft-paste" to form new augmentations after image-level transforming.

## 2.2. Data Augmentations

Data augmentation is an effective solution to the absence of large labeled datasets for semantic segmentation of medical lesions.

*Traditional data augmentation* is the most common method, including rotation, flip, crop, resize, resample, normalization, elastic deformation, etc. nnUnet[62] used these methods to achieve state-of-the-art on several segmentation tasks. Moreover, the change of image intensity such as Gaussian blur[50] and brightness [51] also have been used to achieve data enhancement. Traditional data augmentation is parametric control and has a little computational cost.

*Generative Adversarial Nets (GAN)* provides new solutions for data-hungry. Conversely, people directly synthesize images with GAN, such as Cirillo et al. [23] use 3D-GAN for brain tumor segmentation. Foroozandeh et al.[52] synthesis medical semantic images from mask to image with conditional Generative Adversarial Nets (cGAN), SPADE[53] On the other hand, Cycle-GAN [54]is used for domain transferring [55,56,57,58], in which the labels from the training target domain are not available. Although cGAN adds a condition to the original GAN to guide the generator's generation process, it is still weaker in parametric control than traditional data augmentation. Moreover, GAN requires a lot of training data and skills.

*Transfer learning* is the improvement of learning in a new task by transferring knowledge from a related task that has already been learned, which allows rapid progress and improved performance with few data. ImageNet pretraining [59,60] has been proven effective for performing 2D transfer learning for medical segmentation tasks. Due to an existing a noticeable domain gap between natural images and medical images, Zhou et al.[61] proposed a novel pretraining method that learns from 700k radiographs given no manual annotations to replace the ImageNet pretraining.

*Image mixing* is a class of augmentations that mix the information contained in different images. Zhang et al. [30] are the first to use mixup to improve the robustness of the model. Cutmix[65] is an improved version of the former which randomly cut a rectangular area instant of the whole image. Eaton-Rosen et al.[29] match the crops with the highest foreground amounts and the lowest instant random as a mixup[30]. The data augmentation methods above are usually suitable for the classification and detection tasks, which do not require a highly credible context of object boundary. Moreover, they are still not object-aware.

### 2.3. Copy-patse Augmentations

In contrast, to other data augmentation methods, copy-paste is a stand-out object-aware manner. As for semantic segmentation of natural images, Ghiasi et al.[27] proved that they do not need to model surrounding visual context as [25,26] proposed with sufficient experiments. Poisson blending [64] is a famous classical method of image blending, which has been used to combine Copy-Paste to paste objects naturally by Dwivedi et al.[24]. However, surrounding contexts are essential for medical lesions. As shown in Figure 2, Poisson blending achieves a natural paste by costing the information of medical lesions, which leads to errors in clinical information. Gaussian blurring is not enough to smooth paste-object edges as real medical lesions. Our method compromises the two methods above, which retain lesion information and pastes lesion closer to real images.

### 2.4. Data Imbanlance

Data imbalance is also a severe barrier to accurate semantic segmentation of medical lesions. Especially, the lesions occupy a tiny proportion of pixels in the 3D datasets such as KiTS19[2] BraTS[1], LITS[32]. Focal loss [14] is a popular function loss that puts more focus on hard examples. Salehi et al.[65] proposed the Tversky loss to trade off the penalties between FPs and FNs for image segmentation. In contrast, Data augmentation is a very effective method to balance data, but it has received little attention. Yang et al.[28]proposed TumorCP, which is the first work exploring the "Copy-Paste" design in the medical imaging domain. Our method ensures the correct surrounding context, fully exploring the effect of Copy-Paste augmentation in the medical image segmentation domain.

## 3. Method

### 3.1. Approach Overview

We propose a credible and effective offline data augmentation for semantic segmentation of medical lesions. Figure 3 show the main steps of the Soft-CP augmentation method:

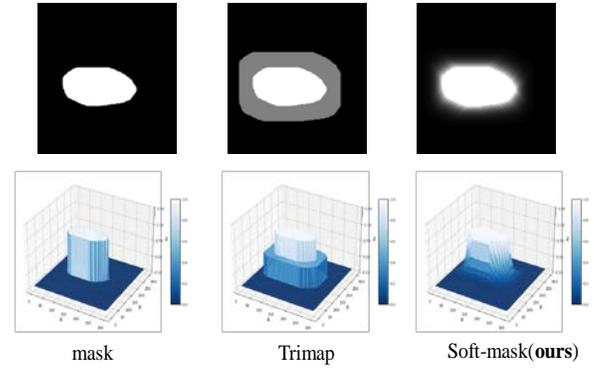

**Figure 4.** Illustration of the mask, Trimap, and Soft-mask with 2D and 3D view.

*Collect lesion and background images:* The main image and paste images are randomly sampled from annotated datasets. Lesion instances are copied from paste images with the method of "Soft-CP." The main image does the image-level transformation to generate background images.

*Transformation:* To improve the variety of augmentations, we do an object-level transformation for lesion instances and an image-level transformation for backgrounds and synthetic images. The probability of each transformation being used is random

### 3.2. Soft-Copy and Soft-Paste (Soft-CP)

**Soft-Copy** is to obtain the information around the lesion by smoothing the edge of the mask, which is different from copying the lesion from the image by directly using the mask and ignoring the information around the lesion. The most similar to Soft-CP is the classical static matting algorithm Trimap. Trimap divides a given image into foreground, background, and unknown region to be solved. The unknown region to be solved contains the boundary of the segmentation instance, which is the key of image matting. In contrast, Soft-CP measures the correlation between edge information and lesions by distance. To be specific, Soft-CP assigns different weights according to the distance of the lesion. e.g., The farther away from the edge of the lesion, the less relevant the environmental information is to the lesion. As Figure 4 shows, to take the binary mask as an example, The unknown region (gray region in Figure 4) pixel value in Trimap is a single value, while the Soft-mask decreases with the distance from the lesion. The Soft-mask multiplies paste image to generate the soft lesion, which contains the lesion and the information around it. This operation is called *Soft-Copy*.

As show in Figure 5. In the implementation, we can control the smoothness of the Soft-mask's edge by the parameter $\alpha$. The value of $\alpha$ is customized based on the dataset. And the soft-mask only needs a simple expansion and corrosion algorithm to generate it.

**Algorithm** soft-mask

**Input**: $M, k_{erode}, k_{dilate}, \alpha$

Firstly erode the mask $M$

$M_{erode} = M$

**while** $i < k_{erode}$ **do**

$\quad M_{erode} = f_{erode}(M_{erode})$

$\quad i = i + 1$

**end while**

Secondly dilate and soften $M_{erode}$

$M_{dilate} = M_{erode}$

**while** $j < k_{dilate}$ **do**

$\quad M_{dilate} = f_{\text{binarization}}(M_{dilate}, threshold = 1e-5)$

$\quad M_{dilate} = f_{dilate}(M_{dilate})$

$\quad M'_{dilate} = M_{dilate} \times \alpha^{j+1}$

$\quad M_{dilate} = (1 - M_{dilate}) M'_{dilate} + M_{dilate}$

**end while**

**Output**: $S = M_{dilate}$

**Figure 5**. **Algorithm** soft-mask calculates the proposed *soft-mask*, $M$ is the binary mask to be softened, $k_{erode}$ and $k_{dilate}$ is the number of iterations for the function of erosion and dilation, $\alpha$ is the softening coefficient and $\alpha \in (0,1)$. $f_x$ represents a function with $x$ capability.(e.g., $f_{erode} = cv2.erode()$ ).Output S is the soft-mask which we wanted.

**Soft-Paste** is to paste the soft-lesion instance generated by Soft-Copy onto the background. The formula (1,2) obtained the synthetic image $I_{syn}$ and the corresponding mask $M_{syn}$. $S$ is the soft-mask which is generated by $M_p$. $M_g$ is the mask of background $I_g$.

$$I_{syn} = I_{soft} + (1 - S) \cdot I_g \quad (1)$$

$$M_{syn} = M_p \cup M_g \quad (2)$$

### 3.3. Transformation

Medical lesions have a certain degree of randomness in morphology, location, size, etc. In this paper, we combine traditional data augmentation with Soft-CP to increase the diversity of the synthetic images. As the Figure3 shows, Object-level transformation guarantees the diversity of the pasted lesions, and image-level transformation guarantees the diversity of the backgrounds and synthetic images.

In the object-level transformation, we only use methods easily controlled by parameters, such as rigid transformation, including scaling, rotation, panning, and flip. Moreover, Soft-CP carries information around the lesion, so it is necessary to prevent overfitting problems. Good experimental results in this paper show that changing image intensity such as Gamma transformation, Gaussian blurring, and Gaussian noise disturbance are good ways to avoid model overfitting, which does not change the clinical judgment of the lesion.

In the image-level transformation, we focus on the shape and pixel intensity of the whole image. With the operation of crop first and then resizing, we cut parts from the image and ensure that the size of the final image is consistent. For changing the image intensity, we still use Gamma transformation and Gaussian transformation.

Moreover, We effectively increase the variety of transformations by random permutations and combinations. Set object-level transformation as an example, rigid transformations: [none, flip, rotation, scaling, panning], intensity transformations: [none, Gamma, Gaussian-N, Gaussian-B], we separately choose two transformations from rigid, and intensity transformation used randomly ordered sampling with replacement. We get an object-level transformation, e.g., [rotation, flip, Gaussian, none].

### 3.4. Where to paste the lesion?

In priors works, Dvornik et al.[25] shows that modeling appropriately the visual context surrounding objects is crucial to place them in the right environment. Fang et al. [26] proposed a location probability map to guide Copy-pasting, which can be placed based on local appearance similarity. However, notice the difference between medical images and natural images. The former has a relatively single background and strict anatomical structure constraints. In this paper, we only need to pay attention to the intersection of the lesion and the reference substance to meet the above two requirements. e.g., on the KiTS19 dataset, the kidney tumor must intersect with the kidney (reference substance), which we can describe the location of the tumor in terms of the number of intersecting pixels as formula (3), indicates the degree of overlap between the kidney tumor and the kidney if it indicates that the augmentation violates anatomical knowledge.

$$S_{lesion} \cap S_{refer} > S_1 \quad (3)$$

Moreover, Under the constraints of the anatomy, the location of the kidney tumor is random. As formula (4) shows, when $S_2 \leq 0$, it indicates that there is no overlap between the two tumors.

$$S_{lesion} \cap S_{refer-lesion} < S_2 \quad (4)$$

By formula (3) and formula (4), we not only ensured the correctness of the anatomical features of the lesions, but also ensured their diversities.

## 4. Experiments

### 4.1. Details

In this paper, to prove the effectiveness of our method, all experiments are based on Unet[7]. The input has been resized as 256×256. The loss function is dice loss[13]. All models are trained with 100 epochs, and the final model we test is the best validation loss set in 100 epochs. The batch size is 8. The initial learning rate is 0.0005, and we use the *CosineAnnealingLR* in PyTorch to update the learning rate with *MAX_STEP=150.* All experiments are conducted on an NVIDIA with 8 12 GB TITAN 2080Ti GPU. And the data of training: validation: testing is approximately equal to 5:1:1. Moreover, in the experiment of data augmentation, the data of real images: synthetic images about 3:1 to 2:1.

### 4.2. Datasets

**BUSI**[35] (Breast Ultrasound Images Dataset) is a dataset for semantic segmentation collected at baseline, including breast ultrasound images among women between 25 and 75 years old. It consists of 780 images with an average size of 500×500 pixels, categorized into normal 133, benign 487, and malignant 210.

**KiTS19**[2] is a 3D collection of multiphase CT imaging, segmentation masks, and comprehensive clinical outcomes for 300 patients who underwent neurectomy for kidney tumors. It randomly selects 210 patients as the training set for the 2019 MICCAI KiTS Kidney Tumor Segmentation Challenge with the most size of 512×512. This data usually serves for benchmarking semantic segmentation models.

**ISIC18**[6] was published by the International Skin Imaging Collaboration (ISIC) as a large-scale dataset of dermoscopy images. For task1 for segmentation, 2594 dermoscopic images with ground truth segmentation masks were provided for training. 100 and 1,000 images were provided for validation and test sets, respectively, without ground truth masks. The size of the skin lesion segmentation dataset ranged from 720×540 to 6708×4439.

**LGG**(lower-grade gliomas)[66] selected brain MR images together with manual FLAIR abnormality segmentation masks. It consists of 3930 images of 110 patients from 5 institutions with lower-grade gliomas from The Cancer Genome Atlas with the size of 256×256.

### 4.3. Evaluation metrics

Evaluation metrics of segmentation accuracy was based on:

i) Dice similarity coefficient (DSC) is a statistic used to gauge the similarity of segmentation and the ground-truth, which is defined as:

$$DSC = \frac{2|X \cap Y|}{|X| + |Y|}$$

where $X$ and $Y$ denote the region segmented by models and the groundtruth.

ii) Accuracy is also used as a statistical measure of the similarity of segmentation and the groundtruth, it is the proportion of correct predictions among the total number of cases examine and is defined as:

$$Accuracy = \frac{TP + TN}{TP + TN + FP + FN}$$

iii) Intersection over Union(IoU) is defind as:

$$IoU = \frac{TP}{TP + FP + FN}$$

where TP = True positive; FP = False positive; TN = True negative; FN = False negative.

### 4.4. Results on the baseline

In order to prove the effectiveness and generality of our method, the datasets we selected include KiTS19[2], BUSI[35], LGG[66], ISIC18[6], and Kvasir-SEG[5]. They cover popular forms of medical imaging such as MRI, Ultrasound, CT, and Optics. The ablation experiment was set up with three different data combinations: real images, real images+ Gaussain-CP images, and real images +Soft-CP images(ours). Moreover, the synthetic images from the Gaussian-CP and Soft-CP are consistent in transformations and numbers in the training process.

As table 3 shows, for BUSI and ISIC18, the more significant pixel proportion of the lesion leads to the smaller gain of the synthetic image for the training result. In particular, it is even worse for added Gaussian-CP to perform than the real images only. For KiTS19 and LGG, the synthetic images from Soft-CP help improve the DSC of the result by 10+%. These experiments demonstrate the generality and effectiveness of our method in data augmentation of the medical lesion, especially in the case of data imbalance.

### 4.5. Results on the low data regime

In this section, we show that Soft-CP is helpful to increase data efficiency. As figure 6 shows, the Copy-Paste augmentation is always helpful for all fractions of KiTS19 [2]. In contrast, Soft -CP is better than Gaussian-CP augmentation. Soft-CP is most helpful in the low data regime (10% of KiTS19), which yields +26.5% DSC over the baseline. Surprisingly, we get a better training model

**Table 3.** We are computing the performance of the model on Gaussian-CP and Soft-CP augmentation for data-efficient for semantic segmentation of medical lesions.

| Dataset | | Combinations | DSC(%) | Accuracy(%) | IoU(%) |
|---|---|---|---|---|---|
| BUSI[35] | Benign | Real | 83.27 | 98.07 | 74.04 |
| | | Real+Guassian-CP | 80.92 | 97.74 | 71.98 |
| | | Real+Soft-CP | 84.78(+1.51) | 98.21(+0.14) | 76.81(+2.77) |
| | Malignant | Real | 61.65 | 89.13 | 49.36 |
| | | Real+Guassian-CP | 61.97 | 90.47 | 49.05 |
| | | Real+Soft-CP | 62.47(+0.82) | 91.62(+2.49) | 49.86(+0.30) |
| KiTS19[2] | Kidney | Real | 85.79 | 99.35 | 75.18 |
| | | Real+Guassian-CP | 85.50 | 99.35 | 74.74 |
| | | Real+Soft-CP | 85.83(+0.04) | 99.38(+0.03) | 75.23(+0.05) |
| | tumor | Real | 64.24 | 99.79 | 51.90 |
| | | Real+Guassian-CP | 69.23 | 99.83 | 57.61 |
| | | Real+Soft-CP | 74.64(+10.40) | 99.86(+0.07) | 62.81(+10.91) |
| ISIC18[6] | | Real | 86.04 | 92.75 | 77.90 |
| | | Real+Guassian-CP | 86.31 | 92.56 | 78.39 |
| | | Real+Soft-CP | 87.70(+1.64) | 93.01(+0.26) | 79.67(+1.77) |
| LGG[66] | | Real | 43.21 | 99.10 | 34.52 |
| | | Real+Guassian-CP | 54.76 | 99.49 | 54.19 |
| | | Real+Soft-CP | 56.63(+13.42) | 99.62(+0.52) | 56.46(+21.94) |

by adding synthetic images to just 40% of the real images rather than 100% real images.

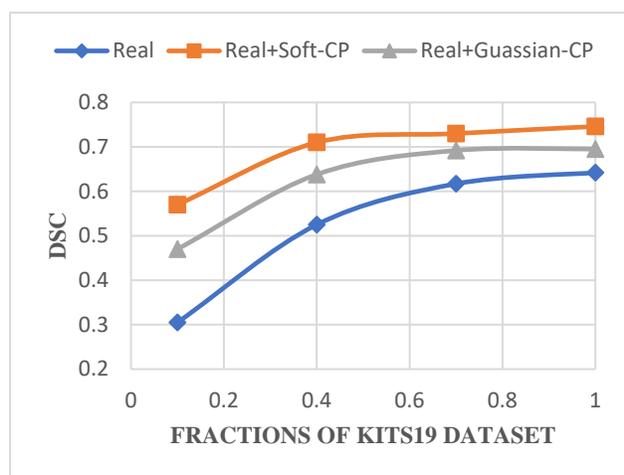

Figure 6. Study of the improvement from Soft-CP and Gaussian-CP to the low data regime.

## 5. Discussion and Future Work

We presented a credible and effective data augmentation for semantics segmentation of medical lesions. Our main contribution is to blend lesions into the background and ensure the correct surrounding information around lesions. Soft-CP is a direct solution to the problem of data scarcity, especially for data imbalance. The experiments proved that our method is suitable for different imaging modalities. And the low data regime combines with our approach can save a lot of storage space and computing costs.

Moreover, offline data augmentation is a controllable method to synthesize the images we need. In future works, we will also try to use it in lesion detection and classification tasks.